\documentstyle[12pt,epsfig]{article}
\begin{document}

\begin{titlepage}
\rightline{May 2013}
\vskip 3cm
\centerline{\Large \bf
Direct detection experiments
}
\vskip 0.4cm
\centerline{\Large \bf
explained with mirror dark matter 
}
\vskip 1.8cm
\centerline{R. Foot\footnote{
E-mail address: rfoot@unimelb.edu.au}}

\vskip 0.5cm
\centerline{\it ARC Centre of Excellence for Particle Physics at the Terascale,}
\centerline{\it School of Physics, University of Melbourne,}
\centerline{\it Victoria 3010 Australia}
\vskip 2cm
\noindent
Recently, the CDMS/Si experiment has observed a low
energy excess of events in their dark matter search. 
In light of this new result we update the mirror dark
matter explanation of the direction detection experiments.
We find that the DAMA, CoGeNT, CRESST-II and CDMS/Si data can be simultaneously
explained by halo $\sim Fe'$ interactions provided that
$v_{rot} \approx 200$ km/s. Other parameter space is also possible.  
Forthcoming experiments, including CDMSlite, CDEX, COUPP, LUX, C-4,... should be
able to further scrutinize mirror dark matter and closely related hidden
sector models in the near future.

\end{titlepage}


The experimental effort to directly detect dark matter has 
been progressing extremely well over the last decade.
Impressive positive results have been reported by DAMA \cite{dama1,dama2}, CoGeNT\cite{cogent,cogent2}, 
CRESST-II\cite{cresst-II}
and now CDMS/Si\cite{cdms-si}. 
Previous work\cite{foot1,foottalk} (see also ref.\cite{footold} for earlier studies) has shown that the
positive results from the DAMA, CoGeNT and CRESST-II 
experiments can be self consistently explained
within the framework of mirror dark matter (for a review see e.g.\cite{review} and references there-in
for a more complete bibliography). 
More generic hidden sector dark matter models are also possible\cite{foot2}.
The purpose of this note is to update the mirror dark matter explanation
in light of the new results from the CDMS/Si experiment.

Recall, mirror dark matter features a hidden sector
which is isomorphic to the ordinary sector. 
That is, interactions are described by the Lagrangian\cite{flv}:
\begin{eqnarray}
{\cal L} = {\cal L}_{SM} (e, \mu, u, d, A_\mu, ...) +
{\cal L}_{SM} (e', \mu', u', d', A'_\mu, ...) + {\cal L}_{mix} \ .
\end{eqnarray}
The theory contains an exact and unbroken parity symmetry: $x \to -x$ 
provided that left and right chiral fields are interchanged in the mirror sector. 
The ${\cal L}_{mix}$ 
part denotes terms coupling the two sectors together. 
Chief among these is
kinetic mixing of the $U(1)_Y$  and $U(1)'_Y$ gauge bosons which is a gauge invariant and renormalizable
interaction\cite{he}.  
This $U(1)$ kinetic mixing implies
also photon-mirror photon kinetic mixing:
\begin{eqnarray}
{\cal L}_{mix} = \frac{\epsilon}{2} F^{\mu \nu} F'_{\mu \nu}
\label{kine}
\end{eqnarray}
where $F_{\mu \nu}$ and $F'_{\mu \nu}$ are the field strength tensors for the photon
and mirror photon respectively. 
One effect of the kinetic mixing interaction is to 
give the mirror electron and mirror proton a 
tiny ordinary electric charge, $\epsilon e$\cite{holdom}. 
This means that
a mirror nucleus, $A'$, with
atomic number $Z'$ and velocity $v$
can Rutherford scatter off an ordinary nucleus, $A$, with
atomic number $Z$. The cross-section for this process is given by 
\begin{eqnarray}
{d\sigma \over dE_R} = {2\pi \epsilon^2 Z^2 Z'^2 \alpha^2 F^2_A F^2_{A'} \over m_A E_R^2 v^2}
\label{cs}
\end{eqnarray}
where 
$F_A$ [$F_{A'}$] is the form factor 
which
takes into account the finite size 
of the nucleus [mirror nucleus]\footnote{Unless otherwise indicated, natural 
units with $\hbar = c = 1$ are used.}.
The Helm form factor\cite{helm,wimp} is used in our numerical work.

The astrophysics and cosmology of kinetically mixed mirror dark matter 
has been discussed in a number of articles e.g. \cite{sph,sil,fc,paolo2,foot69,fstudy}.
A consistent picture appears to be emerging: Mirror dark matter can be the inferred
dark matter in the Universe provided kinetic mixing exists with strength $\epsilon \sim 10^{-9}$.
In this scenario,
dark matter halos in spiral galaxies are presumed to be
composed of mirror particles 
in a pressure supported multi-component plasma
containing $e'$, $H'$, $He'$, $O'$, $Fe'$,...\cite{sph}.  
Such a plasma dissipates energy
due to thermal bremsstrahlung and other processes and this
energy must be replaced.  Studies have shown\cite{sph,fstudy} that ordinary supernovae can supply 
this energy if 
photon-mirror photon kinetic mixing has strength $\epsilon \sim 10^{-9}$ 
and the halo contains a significant mirror metal component ($\stackrel{>}{\sim} 1\%$ by mass).

The mirror metal component can be probed in current direct detection experiments.
The rate depends on the dark matter distribution which is assumed to be Maxwellian
with a temperature, $T$.
This temperature can be roughly estimated from the
hydrostatic equilibrium condition\cite{sph}:
\begin{eqnarray}
T \simeq \frac{1}{2} \bar m v_{rot}^2 
\label{rotx}
\end{eqnarray}
where $v_{rot} \sim 240$ km/s is the galactic rotational velocity and
$\bar m = \sum n_{A'} m_{A'}/\sum n_{A'}$ is the mean mass of the
mirror particles in the halo\footnote{This rough estimate assumed an isothermal
halo. In reality, the temperature is not expected to be constant, but increases
towards the galactic center. However, numerical work indicates\cite{fstudy} that the
temperature at the Sun's location is roughly consistent (i.e. within around 20\%) with the estimate
Eq.(\ref{rotx}).}.
In our numerical work we set
$\bar m \approx 1.1 $ GeV which is suggested by
mirror BBN computations
for $\epsilon \sim 10^{-9}$\cite{paolo2}.  
The halo distribution 
of a mirror nuclei, $A'$, is: 
\begin{eqnarray}
f_{A'}({\textbf{v}},{\textbf{v}}_E) 
 = exp(-E/T) = exp(-\frac{1}{2} m_{A'} {\bf{u}}^2/T) = exp(-{\bf{u}}^2/v_0^2)
\end{eqnarray}
where ${\bf{u}} = {\bf{v}} + {\bf{v}}_E$.
Here ${\bf{v}}$ is the velocity of the halo particles in the Earth's reference frame and ${\bf{v}}_E$ 
is the velocity of the Earth around the galactic center\footnote{We have neglected here the possibility 
of any bulk halo rotation. In the presence of bulk halo motion, ${\bf{v}}_E$ is the velocity
of the Earth with respect to a reference frame where the halo has no bulk motion.} [$\langle v_E \rangle =
v_{rot} + 12$ km/s].
Clearly 
\begin{eqnarray}
v_0[A'] &=& \sqrt{{2T \over m_{A'}}} 
        \simeq  v_{rot} \sqrt{{\bar m \over m_{A'}}}\ . 
\label{v0}
\end{eqnarray} 
Evidently,
the quantity $v_0 [A']$ which characterises the velocity dispersion of the particle $A'$
depends on the mass of the particle. 
This result, along with the recoil energy dependence of the Rutherford scattering
cross-section [$d\sigma/dE_R \propto 1/E_R^2$] 
are important distinguishing features of mirror dark matter  
(and more generic hidden sector models with unbroken $U(1)'$ gauge interactions).

The rate for $A'$ scattering on a target nuclei, $A$, is 
\begin{eqnarray}
{dR \over dE_R} = N_T n_{A'} 
\int^{\infty}_{|{\textbf{v}}| > v_{min}}
{d\sigma \over dE_R}
{f_{A'}({\textbf{v}},{\textbf{v}}_E) 
\over 
v_0^3 \ \pi^{3/2}
} |{\textbf{v}}| d^3 {\textbf{v}} 
\label{55}
\end{eqnarray}
where the integration limit is
$\ v_{min} \ = \ \sqrt{ (m_{A} + m_{A'})^2 E_R/2 m_{A} m^2_{A'} }$\ .
In Eq.(\ref{55}), $N_T$ is the number of target nuclei and 
$n_{A'} = \rho_{dm} \xi_{A'}/m_{A'}$ 
is the number density of the halo $A'$ particles. 
[$\rho_{dm} = 0.3 \  {\rm GeV/cm}^3$ and $\xi_{A'}$ is the halo mass fraction
of species $A'$].
The integral, Eq.(\ref{55}), can be expressed in terms
of error functions and numerically solved.


Detector resolution effects can be included by convolving the rate
with the appropriate Gaussian distribution.
The relevant rates for the DAMA, CoGeNT, CRESST-II and CDMS/Si experiments   
can then be computed and compared with the data.
Note that the $H', \ He'$ halo components are too light to give significant signal contributions
due to exponential kinematic suppression. For these experiments only heavier `metal' components can give an observable 
signal above the detector
energy thresholds.
For simplicity we assume that the rate in each experiment is dominated by the interactions 
of a single such metal component, $A'$. 
Naturally this is only an approximation, however it can 
be a reasonable one given the fairly narrow energy range probed in the
experiments [the signal regions are mainly:
2-4 keVee (DAMA), 0.5-1 keVee (CoGeNT),  12-14 keV (CRESST-II) and 7-13 keV (CDMS/Si)].
With this approximation, the
scattering rate depends on the parameters $m_{A'},\ \epsilon \sqrt{\xi_{A'}}$
and also $v_{rot}$. 


The CDMS/Si experiment has recorded three dark matter
candidate events in a 140.2 kg-day exposure of an array of silicon detectors\cite{cdms-si}.
The nominal recoil energies of these three events are 8.2 keV, 9.5 keV and 12.3 keV.
Since the number of events is low we cannot perform a $\chi^2$ analysis.
Instead, we use the extended maximum likelihood formalism\cite{barlow} to construct
the likelihood function.
This has the form:
\begin{eqnarray}
{\cal L}({\bf p}) = \left[ \Pi_{i} {dn (E_R^i) \over dE_R} \right] exp[-{\cal N({\bf p})}]
\end{eqnarray}
where the vector ${\bf p}$ denotes the unknown parameters.
Here, $dn (E_R^i)/dE_R$ 
is the expected event rate evaluated at the recoil energy for each of the three observed events, $i=1,...,3,$ and
$\cal N({\bf p})$ is the total number of expected events in the
acceptance recoil energy region:
\begin{eqnarray}
{\cal N({\bf p})} = \int {dn \over dE_R} \ dE_R
\ .
\end{eqnarray}
The expected event rate, $dn/dE_R$, is given by the rate $dR/dE_R$ 
convolved with a Gaussian to take into account the resolution\footnote{ 
In the absence of resolution measurements, we take $\sigma_{res} = 0.1$ keV.}
and 
multiplied by the detection efficiency, $\epsilon_f (E_R)$ (obtained from 
figure 1 of ref.\cite{cdms-si}).

According to the CDMS paper\cite{cdms-si}, there are indications that the recoil energy calibration is 
likely around 10\% higher than nominally used, with some uncertainty. We therefore scaled the energies
up by a factor, $f=1.1$ and considered an energy calibration uncertainty of $\pm 10\%$, i.e. $f = 1.1 \pm 0.1$. 
For each value of $m_{A'}, \ \epsilon \sqrt{\xi_{A'}}$ we have maximized ${\cal L}$ over this
range of $f$, to give profile likelihood function, ${\cal L}_P$  \footnote{
In our numerical work we let $A', \ Z'$ have non-integer values, with $Z' = A'/2$, except
when we specifically consider $A' = Fe'$, in which case we use $Z' = 26$, $A' = 55.8$.}.
The favoured region for the parameters $m_{A'},\ \epsilon \sqrt{\xi_{A'}}$
is then determined by the condition 
\begin{eqnarray}
ln \ {\cal L}_P \ge ln \ {\cal L}_{P max} - \Delta \ ln \ {\cal L}_P
\ .
\end{eqnarray}
We fix
$2\Delta \ ln \ {\cal L}_p = 5.99 $ which corresponds to 95\% C.L. for 2 parameters\cite{revppg}. 
In our scan of parameter space we constrain $m_{A'} \le m_{Fe'} \simeq 55.8 m_p$ ($m_p$ is the
proton mass).
Note that we neglect backgrounds in this analysis, since the known backgrounds in the
energy region of interest, $E_{threshold} \le E_R \le 20$ keV, is much less than 1 event. 

The analysis of DAMA, CoGeNT and CRESST-II is analogous to our earlier study\cite{foot1}.
As described there, we analyse the DAMA annual modulation signal obtained from the 1.17 ton-year exposure\cite{dama2}.
We consider 12 bins of width 0.5 keVee\footnote{
The unit keVee is the electron equivalent energy, which is related to nuclear recoil
energy via keVee = keV$/q$, where $q < 1$ is the quenching factor.}
from 2 keVee - 8 keVee and evaluate the theoretical annual modulation
signal as a function of $m_{A'}\ \epsilon \sqrt{\xi_{A'}}$, taking into account
detector resolution effects. We introduce a $\chi^2$ function in the usual way:
\begin{eqnarray}
\chi^2 (m_{A'}, \epsilon \sqrt{\xi_{A'}}) = \sum \left[ {R_i - data_i \over \delta data_i}\right]^2
\ .
\end{eqnarray}
We then minimize $\chi^2$ over quenching factor uncertainty which we take as: $q_{Na} = 0.28 \pm 0.08$ and
$q_I = 0.12 \pm 0.08$. In our analysis we neglect the possibility of channeling\cite{gelmini}\footnote{
There are recent indications that the DAMA quenching factors might be smaller than the considered range\cite{collarq}, 
and some other indications
that the DAMA quenching factors might be larger\cite{tret}. Additionally, 
a few percent channeling fraction for iodine (and also sodium if there are lighter more abundant halo
components) can be important which can significantly lower the DAMA favoured region. In view of these uncertainties,
the DAMA favoured region should be considered as a rough guide only.}.
For CoGeNT, we consider the  most recent
data, stripped of known background components, and corrected for surface event contamination
and overall detection efficiency\cite{cogent2}. 
This data is separated into
15 bins of width 0.1 keVee over the energy range 0.5 - 2 keVee. The resulting 
$\chi^2$ is minimized over the germanium quenching factor uncertainty which we take as: $q_{Ge} = 0.21 \pm 0.04$
and a constant background contribution.
For CRESST-II we bin the data into 5 bins with keV energy ranges of 10.2-13, 13-16, 16-19, 19-25, 25-40.
No energy calibration uncertainty is considered for CRESST-II.
For each data set, we
give $95\%$ C.L. favoured regions [$\chi^2 \le \chi^2_{min} + \Delta \chi^2$, with
$\Delta \chi^2 = 5.99$]\footnote{
An analysis\cite{cf} of the CDMS/Ge low energy data\cite{cdmsgelow} 
strongly supports [5.7 $\sigma$ C.L.] a family of low energy events in the 
nuclear recoil band.
Although we don't specifically include the CDMS/Ge data in our
analysis here, the study\cite{tough} indicates that this data
is compatible
with the overlapping region of parameter space found in figure 1.}.

\begin{table}
\centering
\begin{tabular}{c c c c c}
\hline\hline
$v_{rot}$ [km/s]   & CDMS & CoGeNT & DAMA & CRESST-II\\
               &  &   
$\chi^2$ (min)/d.o.f.  &  
$\chi^2$ (min)/d.o.f.  & 
$\chi^2$ (min)/d.o.f.  \\
& best fit param. &  best fit param. &
 best fit param. &  best fit param. \\
\hline
&  & & & \\
200 &      & 9.7/12 & 5.7/10 & 2.6/3 \\
    & ${m_{A'} \over m_p}$ = 55.8 & ${m_{A'} \over m_p}$ = 39.0 & ${m_{A'} \over m_p}$ = 55.8 & 
    ${m_{A'} \over m_p}$ = 55.8 \\ 
    & $\frac{\epsilon \sqrt{\xi_{A'}}}{10^{-10}} = 0.93$ & ${\epsilon \sqrt{\xi_{A'}} \over 10^{-10}} = 2.5$ & 
    ${\epsilon \sqrt{\xi_{A'}} \over 10^{-10}} = 2.5$ & ${\epsilon \sqrt{\xi_{A'}} \over 10^{-10}} = 2.7$ \\
&  & & & \\
\hline
&  & & & \\
240 &      & 9.9/12 & 5.0/10 & 0.3/3 \\
    & ${m_{A'} \over m_p}$ = 37.0 & ${m_{A'} \over m_p}$ = 31.0 & ${m_{A'} \over m_p}$ = 45.2 & 
    ${m_{A'} \over m_p}$ = 55.8 \\ 
    & ${\epsilon \sqrt{\xi_{A'}} \over 10^{-10}} = 1.2$ & ${\epsilon \sqrt{\xi_{A'}} \over 10^{-10}} = 3.1$ & 
    ${\epsilon \sqrt{\xi_{A'}} \over 10^{-10}} = 3.6$ & ${\epsilon \sqrt{\xi_{A'}} \over 10^{-10}} = 1.7$  
\\
&  & & & \\
\hline
&  & & & \\
280 &      & 10.1/12 & 5.2/10 & 0.2/3 \\
    & ${m_{A'} \over m_p}$ = 25.5 & ${m_{A'} \over m_p}$ = 25.0 & ${m_{A'} \over m_p}$ = 37.7 & 
    ${m_{A'} \over m_p}$ = 36.0 \\ 
    & ${\epsilon \sqrt{\xi_{A'}} \over 10^{-10}} = 1.6$ & ${\epsilon \sqrt{\xi_{A'}} \over 10^{-10}} = 3.6$ & 
    ${\epsilon \sqrt{\xi_{A'}} \over 10^{-10}} = 4.7$ & ${\epsilon \sqrt{\xi_{A'}} \over 10^{-10}} = 2.4$  
\\
&  & & & \\
\hline\hline
\end{tabular}
\caption{Summary of $\chi^2 (min)$ and best fit parameters for the relevant data sets from the
CDMS, CoGeNT, DAMA and CRESST-II experiments.}
\end{table}

In table 1 we summarize the $\chi^2$ minimum values and best fit parameters from each experiment for three representative 
values of $v_{rot}$.
In figure 1 we plot 
the favoured region of parameter space for each experiment, for these same $v_{rot}$ values.
\vskip 0.4cm
\centerline{\epsfig{file=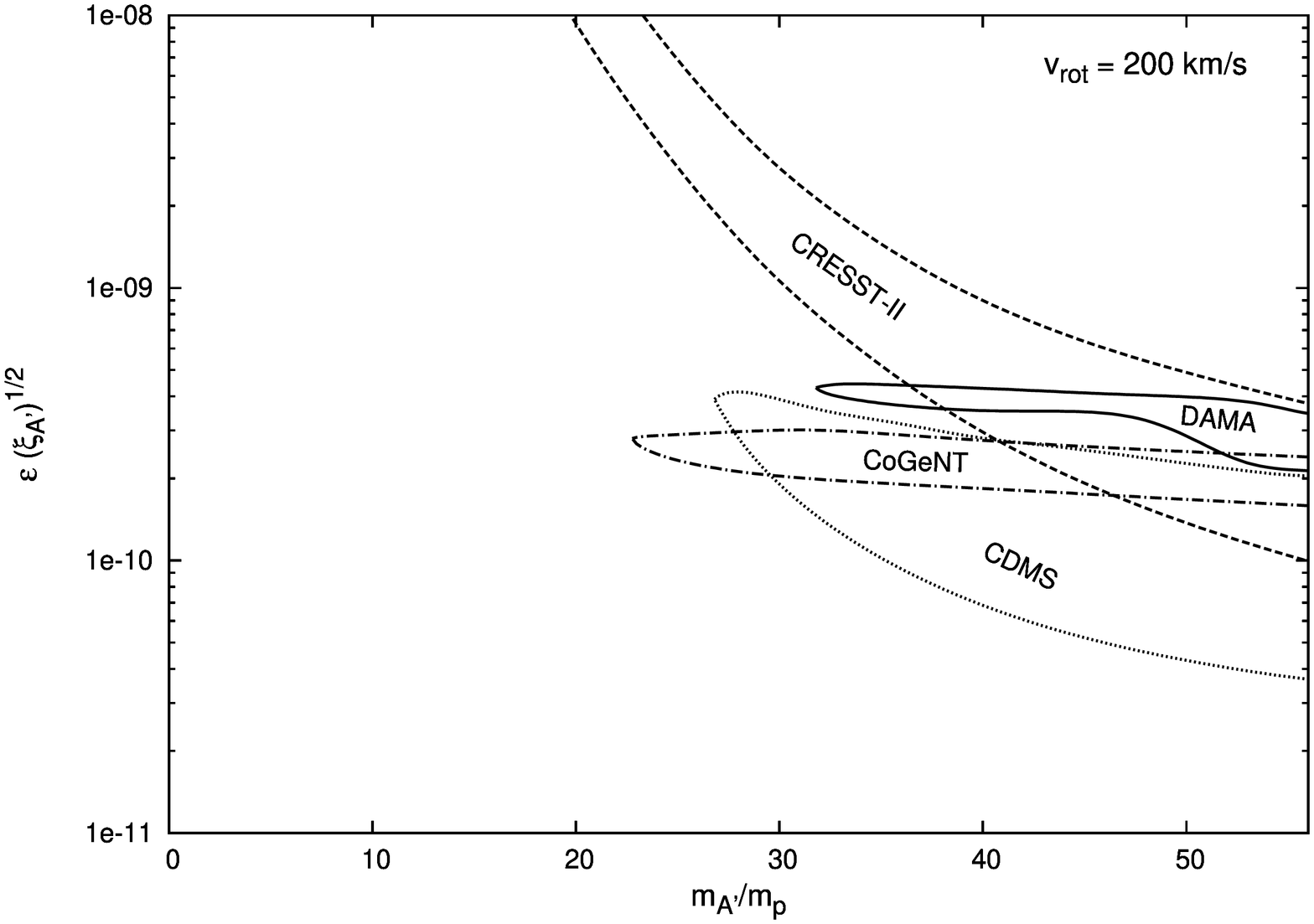,angle=270,width=14.0cm}}
\vskip 0.3cm
\noindent
{\small
Fig 1a: DAMA (solid lines), CoGeNT (dashed-dotted lines), 
CRESST-II (dashed lines) and CDMS/Si (dotted lines) favoured regions of 
parameter space [95\% C.L.] in the mirror dark matter model for
$v_{rot} = 200$ km/s.
}
\vskip 0.5cm
\centerline{\epsfig{file=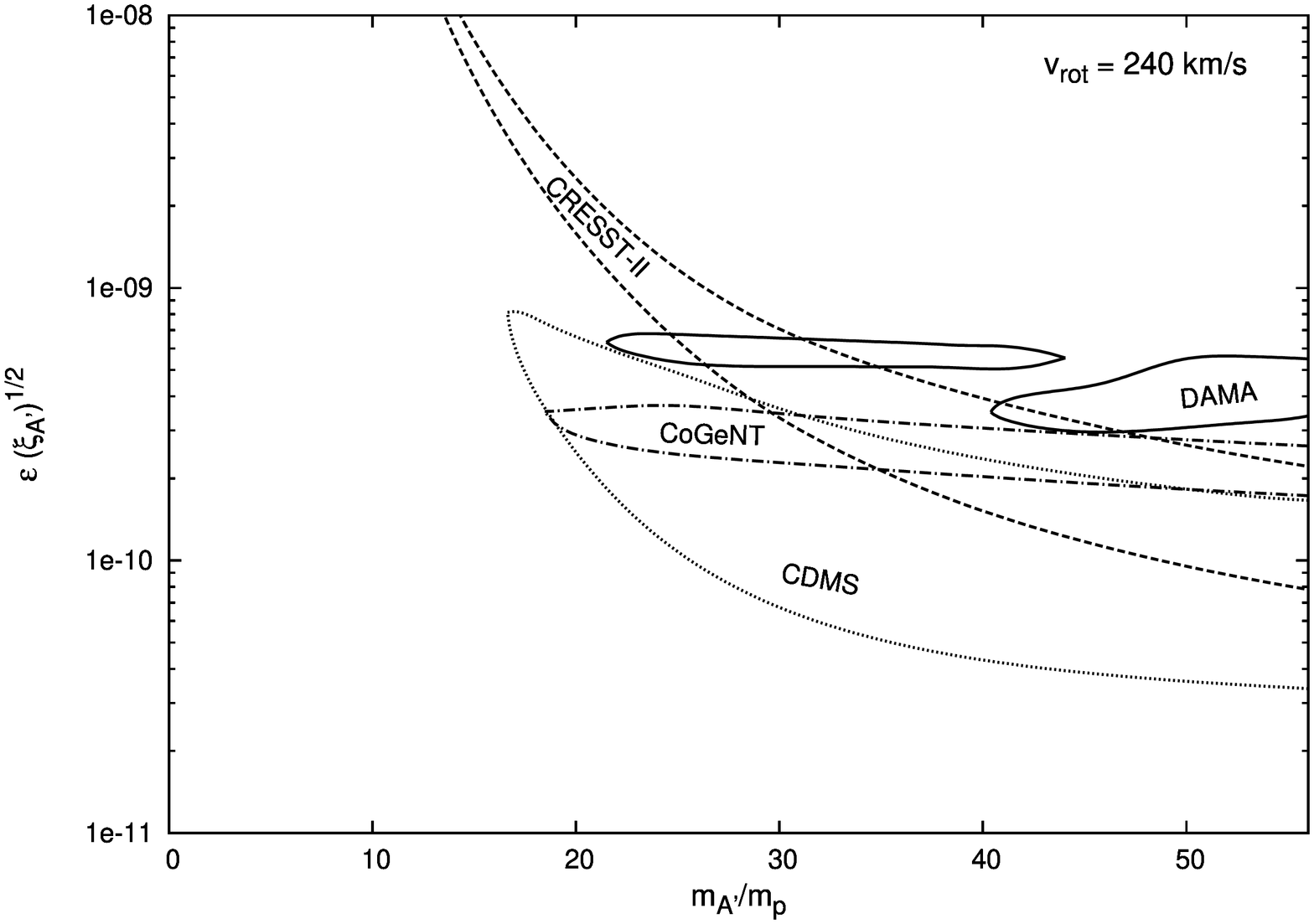,angle=270,width=14.0cm}}
\vskip 0.2cm
\noindent
{\small
Fig 1b: Same as figure 1a, except 
$v_{rot} = 240$ km/s.
}
\vskip 0.5cm
\centerline{\epsfig{file=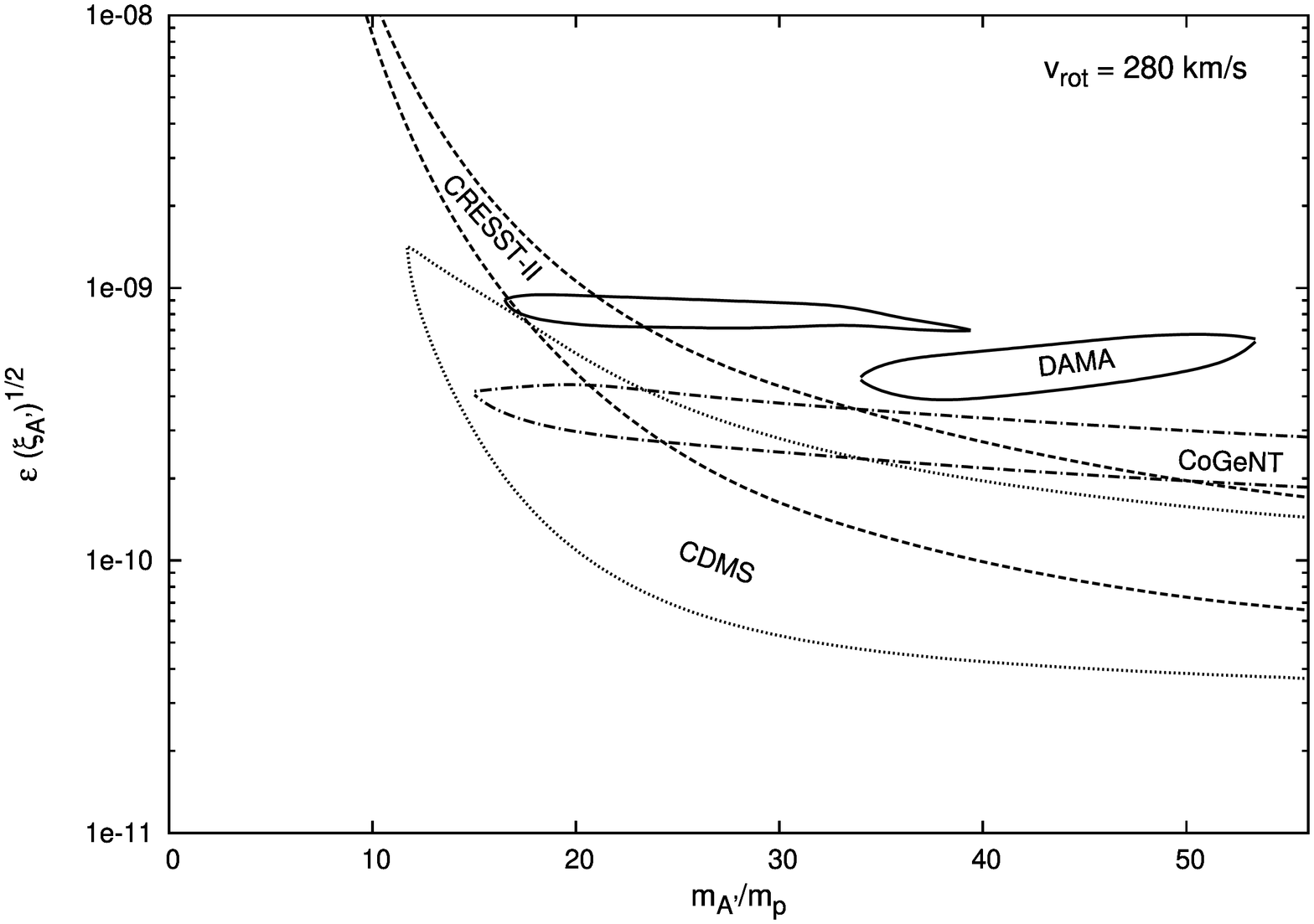,angle=270,width=14.0cm}}
\vskip 0.3cm
\noindent
{\small
Fig 1c: Same as figure 1a, except 
$v_{rot} = 280$ km/s.
}
\vskip 0.9cm
Figure 1 indicates a substantial region of parameter space where
all four experiments can be explained within this theoretical framework.
Although this figure suggests that the experiments favour $A' \sim Fe'$, $v_{rot} \approx 200$ km/s,
the potential uncertainties cannot exclude other parameter space, with lighter $A'$ components and
higher $v_{rot}$ values.

\begin{table}
\centering
\begin{tabular}{c c c c c}
\hline\hline
$m_{A'}/m_p$   & CDMS & CoGeNT & DAMA & CRESST-II\\
$Z'$ &  &   
$\chi^2$ (min)/d.o.f.  &  
$\chi^2$ (min)/d.o.f.  & 
$\chi^2$ (min)/d.o.f.  \\
& best fit param. &  best fit param. &
 best fit param. &  best fit param. \\
\hline
&  & & & \\
55.8 &      & 9.3/12 & 5.5/10 & 0.3/3 \\
26 & $v_{rot}$ = 205 km/s & 
$v_{rot}$ = 150 km/s & 
$v_{rot}$ = 210 km/s & 
$v_{rot}$ = 250 km/s  
\\
    & $\frac{\epsilon \sqrt{\xi_{A'}}}{10^{-10}} = 0.96$ & ${\epsilon \sqrt{\xi_{A'}} \over 10^{-10}} = 1.9$ & 
    ${\epsilon \sqrt{\xi_{A'}} \over 10^{-10}} = 3.1$ & ${\epsilon \sqrt{\xi_{A'}} \over 10^{-10}} = 1.7$ \\
&  & & & \\
\hline
&  & & & \\
28.1 &      & 9.8/12 & 9.3/10 & 0.3/3 \\
14 & $v_{rot}$ = 270 km/s & 
$v_{rot}$ = 210 km/s & 
$v_{rot}$ = 280 km/s & 
$v_{rot}$ = 300 km/s  
\\
    & $\frac{\epsilon \sqrt{\xi_{A'}}}{10^{-10}} = 1.5$ & ${\epsilon \sqrt{\xi_{A'}} \over 10^{-10}} = 2.5$ & 
    ${\epsilon \sqrt{\xi_{A'}} \over 10^{-10}} = 8.1$ & ${\epsilon \sqrt{\xi_{A'}} \over 10^{-10}} = 3.1$ \\
&  & & & \\
\hline
&  & & & \\
16.0 &      & 11.8/12 & 7.6/10 & 3.1/3 \\
8 & $v_{rot}$ = 300 km/s & 
$v_{rot}$ = 300 km/s & 
$v_{rot}$ = 300 km/s & 
$v_{rot}$ = 300 km/s  
\\
    & $\frac{\epsilon \sqrt{\xi_{A'}}}{10^{-10}} = 3.5$ & ${\epsilon \sqrt{\xi_{A'}} \over 10^{-10}} = 3.9$ & 
    ${\epsilon \sqrt{\xi_{A'}} \over 10^{-10}} = 10.1$ & ${\epsilon \sqrt{\xi_{A'}} \over 10^{-10}} = 11.0$ \\
&  & & & \\
\hline\hline
\end{tabular}
\caption{Summary of $\chi^2 (min)$ and best fit parameters for the relevant data sets from the
CDMS, CoGeNT, DAMA and CRESST-II experiments.}
\end{table}


Instead of fixing, $v_{rot}$ and varying $m_{A'}\ \epsilon \sqrt{\xi_{A'}}$
we also consider fixing $A'$ and treating $v_{rot}, \ \epsilon \sqrt{\xi_{A'}}$
as free parameters (subject to the mild constraint, $150 \le v_{rot} [{\rm km/s}] \le 300$).
In table 2 we summarize the $\chi^2$ minimum and best fit points for three such fixed $A'$ choices.
For each of these choices we have 
plotted the favoured region of parameter space in figure 2.
\vskip 0.5cm
\centerline{\epsfig{file=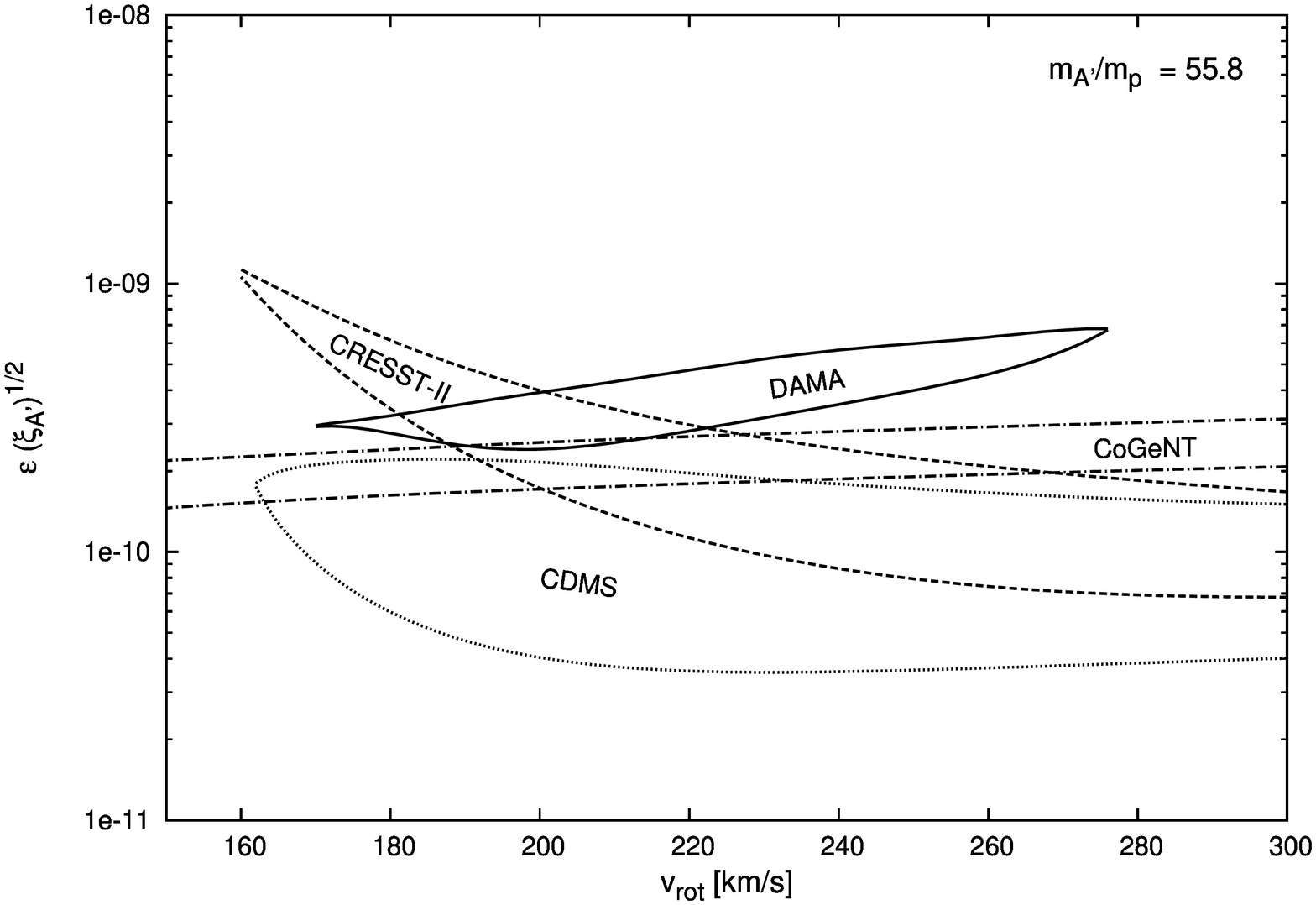,angle=270,width=14.0cm}}
\vskip 0.3cm
\noindent
{\small
Fig 2a: DAMA (solid lines) , CoGeNT (dashed-dotted lines), CRESST-II (dashed lines) and 
CDMS/Si (dotted lines) favoured regions
of parameter space [95\% C.L.] for $A' = Fe'$.
}
\vskip 0.5cm
\centerline{\epsfig{file=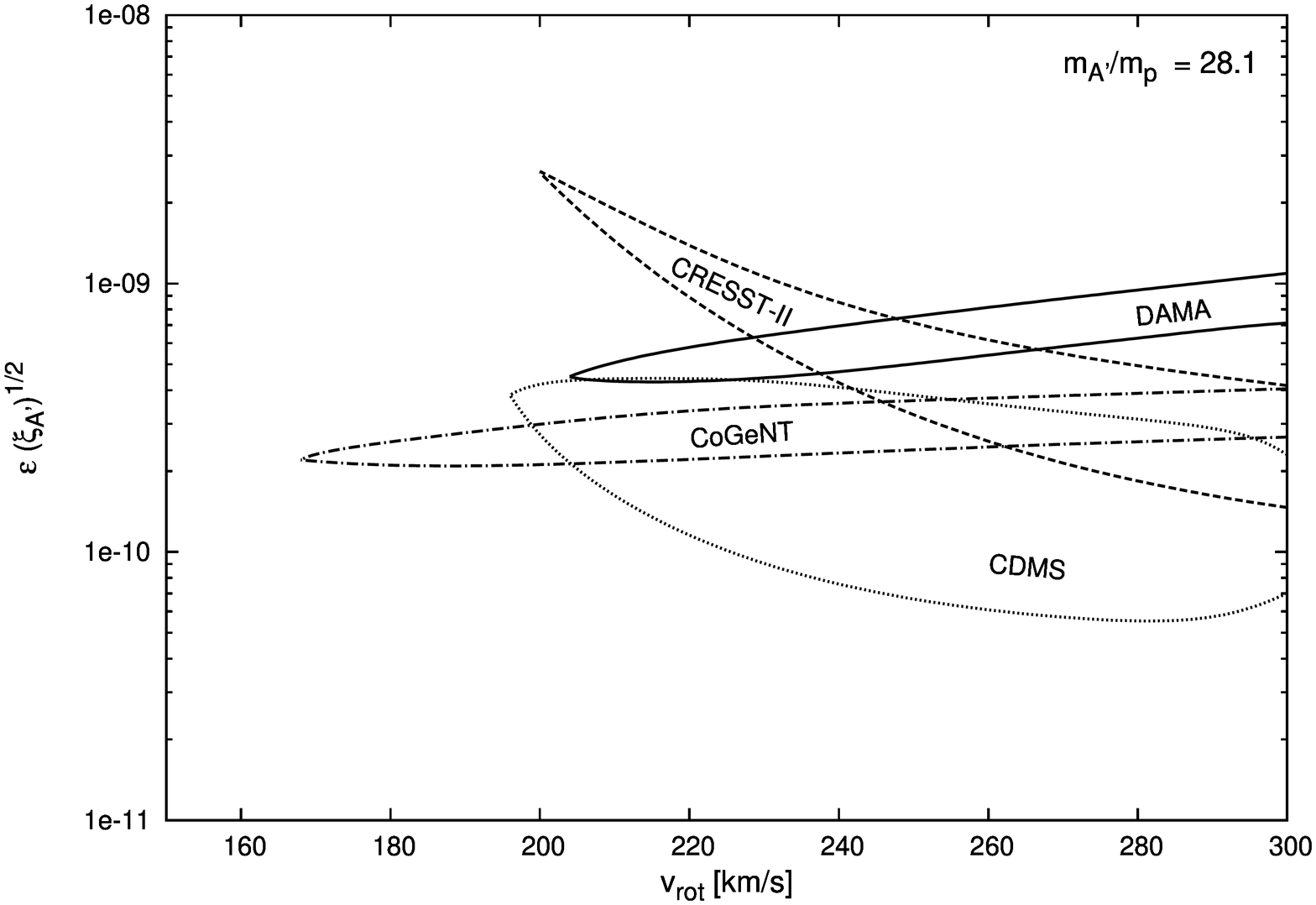,angle=270,width=14.0cm}}
\vskip 0.3cm
\noindent
{\small
Fig 2b: Same as figure 2a except for $A' = Si'$.
}
\vskip 0.5cm
\centerline{\epsfig{file=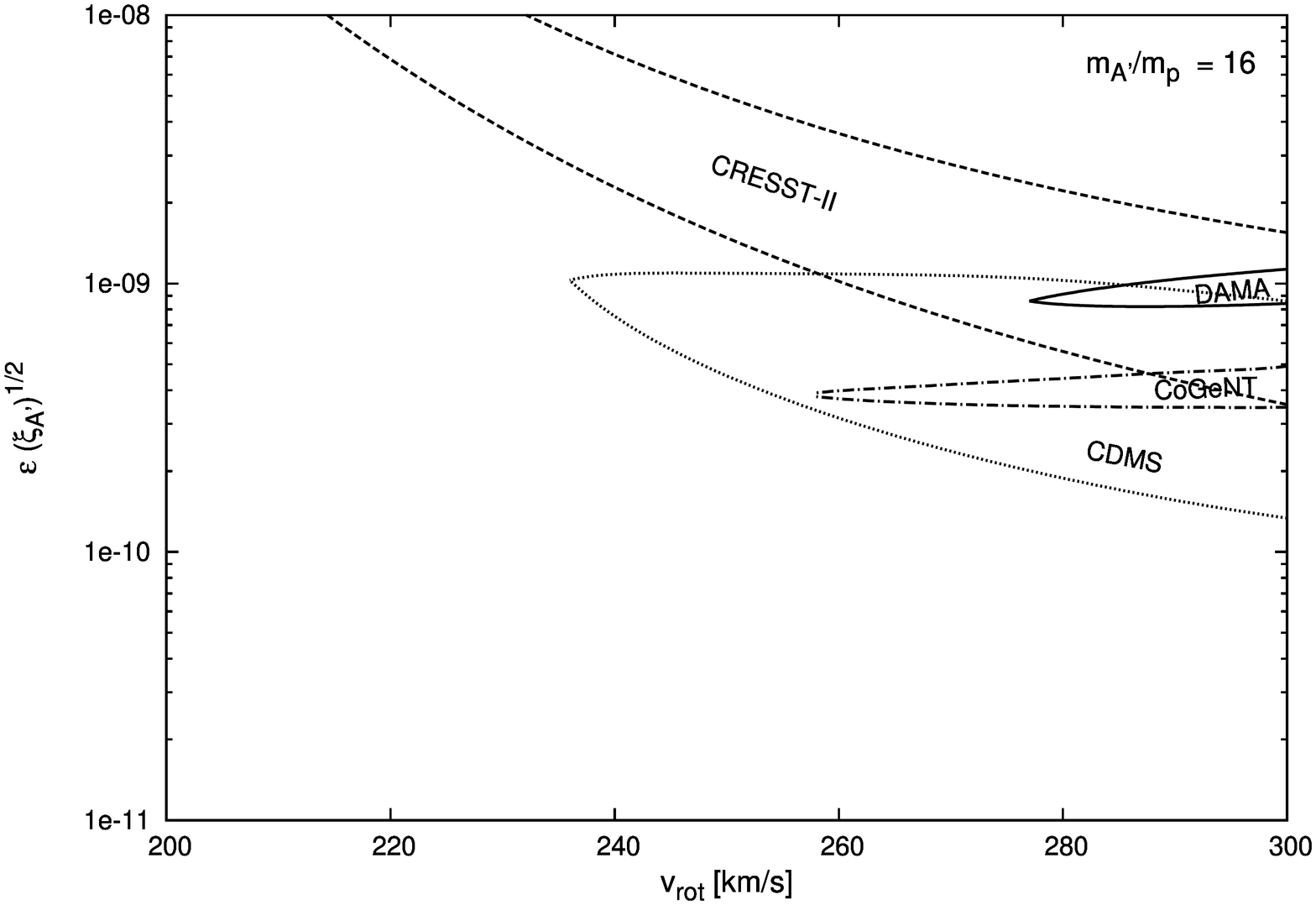,angle=270,width=14.0cm}}
\vskip 0.3cm
\noindent
{\small
Fig 2c: Same as figure 2a except for $A' = O'$.
}
\vskip 1.1cm

If $v_{rot} \sim 200$ km/s and $A' \sim Fe'$ is indeed the component being
detected by DAMA, CoGeNT, CRESST-II and CDMS/Si then we expect lower threshold experiments 
to potentially detect lighter components such as $Si'$ and $O'$ (in addition to $Fe'$). 
TEXONO, CDMSlite, C-4, etc might be sensitive to these components.
Higher threshold experiments such as XENON100, XENON1T, LUX, XMAS, Edelweiss, etc
may only be able to detect the tail of the $Fe'$ halo velocity distribution
This would be manifested
as a bunch of events close to threshold. The two events recorded by XENON100\cite{xenon} and four events
by Edelweiss\cite{edelweiss} near threshold could be an interesting hint of such an observation.

One might have anticipated XENON100  
to have seen a few dozen or so events instead of only 2 events\cite{foot1}. Thus,
the favoured region of parameter space identified in figures 1,2 has significant tension
with the current results 
of the XENON100 experiment\cite{xenon}. The amount of tension, though,  
depends sensitively on the systematic uncertainty one
assigns for XENON's energy calibration (for recent discussions, see e.g.\cite{collarg}). 
For instance the level of tension reduces to zero if the threshold energy
is a factor of $\sim$ 1.5-2.0 times higher than the value used by XENON100.
Recall the XENON100 experiment has no low energy calibration, but extrapolates the energy scale
from measurements of 122 keV X-rays.
Importantly, the XENON100 collaboration has plans to check the 
calibration of their detector in the near future\cite{xenontalk} which might help clarify this situation.
 
Ultimately these very sensitive but higher threshold experiments might also be able to 
probe the (expected) rare $\sim Pb'$ component. We estimate
that the current XENON100 limit on such a  
component is $\xi_{Pb'}/\xi_{Fe'} \stackrel{<}{\sim} 10^{-2}$ at 95\% C.L. if
$\epsilon \sqrt{\xi_{Fe'}} \approx 2 \times 10^{-10}$.
On the other hand, if $v_{rot} \sim 240-280$ km/s so that $A' \sim Si'$ or $O'$ is
being detected by DAMA, CoGeNT, CRESST-II and CDMS/Si then XENON100, XENON1T, LUX, XMAS, Edelweiss etc
would be expected to probe the $Fe'$ component.
We estimate the current limit on $Fe'$ in this case is 
$\xi_{Fe'}/\xi_{Si'} \stackrel{<}{\sim} 10^{-2}$
if $\epsilon \sqrt{\xi_{Si'}} \approx 3\times 10^{-10}, \ v_{rot} = 240 $ km/s and 
$\xi_{Fe'}/\xi_{O'} \stackrel{<}{\sim} 10^{-2}$
if
$\epsilon \sqrt{\xi_{O'}} \approx 4 \times 10^{-10}, \ v_{rot} = 280 $ km/s. 


To conclude,
the DAMA, CoGeNT, CRESST-II and CDMS/Si results
have been examined 
in the context of the mirror dark matter candidate.
In this framework
dark matter consists of a spectrum of mirror particles of known masses: $e'$, $H'$, $He'$,
$O'$, $Fe'$, ... \ .
We find that this theory can simultaneously explain 
the data from each experiment by
$A' \sim  Fe'$ interactions
if $\epsilon \sqrt{\xi_{Fe'}} \approx 2\times 10^{-10}$ and 
$v_{rot} \sim 200$ km/s.
Other parameter regions, and also, more generic hidden sector
dark matter models are also possible.
Further direct tests of this dark matter framework 
are expected in the near future from higher precision
experiments such as C-4, CDEX, 
superCDMS, COUPP, LUX, PandaX etc. These and other experiments should provide
further scrutiny of mirror dark matter  
and closely related hidden sector models in the near future.

\vskip 1.2cm
\noindent
{\large \bf Acknowledgments}

\vskip 0.1cm
\noindent
This work was supported by the Australian Research Council.

\end{document}